\documentclass{mem}
\usepackage{natbib}\usepackage{txfonts}\usepackage{balance}
\usepackage{graphicx}
\usepackage{trfsigns}
\usepackage{txfonts}
\usepackage[a4paper]{hyperref}
\idline{79}{3}

\def\acena{${\alpha}$~{\it Cen}~A}

\def\m2s2{m$^{2}$~s$^{-2}$} 
\def\msol{M${_\odot}$}             

\def\thetav{{\mathbf{\theta}}}
\def\Xv{\mathbf{X}}
\def\stage{t_{\star}}

\begin{document}

\title{Estimation of stellar parameters using Monte Carlo Markov Chains.
}

   \subtitle{}

\author{
M. \,Bazot\inst{1,2}, 
S. \,Bourguignon\inst{3,4} 
\and J. \, Christensen-Dalsgaard\inst{2}
          }

  \offprints{M.~Bazot}

\institute{
Centro de Astrofisica da Universidad do Porto; Rua das Estrelas, 4150-762 Porto, Portugal,
\email{bazot@astro.up.pt}
\and
Institut for Fysik og Astronomi, Aarhus Universitet, DK-8000 Aarhus C., Danmark
\and
Centre Ifremer de Brest, 29280 Plouzan\'e, France
\and
Laboratoire d'Astrophysique de Toulouse-Tarbes, Observatoire Midi-Pyr\'en\'ees, 31400 Toulouse, France
}

\authorrunning{Bazot}

\titlerunning{Estimation of stellar parameters using MCMC}

\abstract{
We apply Monte Carlo Markov Chain methods to the stellar parameter estimation problem. This technique is useful when dealing with non-linear models and allows to derive realistic error bars on the inferred parameters. We give the first results obtained for {\acena}.
\keywords{Stars: fundamental parameters -- Stars: individual: HD128620 -- Methods: statistical}
}
\maketitle{}

\section{Introduction}

   Estimation of the parameters of stellar models is a challenging task. The increasing precision achieved in the determination of atmospheric parameters, coupled to the fast growing amount of available seismic and interferometric data, has allowed to improve the constraints on stellar models.

 However, it is also important to improve the methods for stellar parameter estimation. Stellar models are known to be highly non-linear. Therefore, optimisation methods may fail in computing the optimum parameters. Moreover, it is usually difficult to associate a confidence level to the inferred stellar parameters. 

We consider here the use of Monte Carlo Markov Chain (MCMC) algorithms to address this problem. They offer the advantage of being efficient for non-linear models and allow, when combined with statistical inference, to estimate jointly the parameters and associated confidence levels.

\section{Principle of the MCMC algorithm}
\begin{table*}[t!]
\begin{center}
\caption{Results of the MCMC simulations for {\acena}. For each run the first row gives the mean value and the corresponding 1-$\sigma$ error bar. The second row gives the parameter value estimated from the maximum {\it a posteriori} (MAP, in brackets). }
\label{estim}
\begin{tabular}{lcccc}
\hline
Run \#& $M$ ({\msol})&$\sigma_M$ ({\msol})& $\stage$ (Gyr) & $\sigma_{\stage}$ (Gyr)\\
\hline
1& 1.122 & 0.019 & 5.59 & 1.13\\
&(1.119)&&(5.80)&\\
2& 1.107 & 0.007 & 6.52 & 0.45\\
&(1.106)&&(6.55)&\\
3& 1.117 & 0.014 & 5.91 & 0.51\\
&(1.116)&&(5.97)&\\
\hline
\end{tabular}
\end{center}
\end{table*}
Let $\thetav$ and $\Xv$ be the vectors collecting respectively the model parameters and the observational data. Our probabilistic approach aims at estimating the posterior probability distribution (PPD) of the model parameters, conditional on the available data for the star, which is given by Bayes's formula :
\begin{equation}\label{bayes}
\pi(\thetav|\Xv) = \frac{f(\Xv|\thetav)\pi(\thetav)}{K},
\end{equation}
with $\pi(\thetav|\Xv)$ the PPD, $f(\Xv|\thetav)$ the likelihood, $\pi(\thetav)$ a given prior density function for the model parameters and $K$ a normalisation constant. Assuming that the observations are independent, we used a Gaussian likelihood:
\begin{equation}{\label{likeli}}
f(\Xv|\thetav) \propto \exp \left( - \frac{1}{2}\displaystyle\sum_{i=1}^{N} \left[ \frac{X_i^{th}(\thetav)-X_i}{\sigma_i} \right]^2 \right),
\end{equation}
with $X_i^{\mathrm{th}}(\thetav)$ the output of the model corresponding to the $i$-th component of $\Xv$, $\sigma_i$ the standard deviation, chosen as the $1\sigma$-error bar on the measurement and $N$ the number of measurements.

The idea behind MCMC algorithms is to generate samples distributed according to a distribution of interest, here $\pi(\thetav | \Xv)$. Since these distributions are usually complex, samples are generated using simpler ones, called instrumental distributions. The useful property of an MCMC algorithm is that when it reaches convergence it generates sets of parameters according to the target PPD.

 The most general form of MCMC algorithms, the one we used in this work, is the Metropolis-Hastings algorithm \citep{Metropolis53}. At iteration $t$, a new candidate $\theta^\star$ is sampled from a given instrumental distribution $q(\theta^\star|\theta^{t-1})$, and is chosen as $\theta^t = \theta^\star$ with some acceptance probability depending on $q$, $\pi$, $\theta^{t-1}$ and $\theta^\star$. Distribution $q$ has to be chosen carefully, so that it can cover the whole parameter space and also represent the true posterior distribution, in order to achieve a satisfactory acceptance rate. To this end, we chose a mixture of three distributions: a uniform and a Gaussian distribution with large variance allow to scan the entire parameter space, while a Gaussian distribution with small variance is used for quick local refinements.

\section{A test case: $\alpha$ Cen A}

  Since it is the closest star to the Solar System, {\acena} has been extensively observed. Effective temperature and luminosity, $T_{\mathrm{eff}}=5810\pm50$~K  and $L/L_{\odot}=1.522\pm0.030$ \citep{Eggenberger04}, radius, $R/R_{\odot}=1.224\pm0.003$ \citep{Kervella03} and seismic data are available and, because it is part of a binary, so is its mass, $M/M_{\odot} = 1.105\pm0.007$ \citep{Pourbaix02}. In the following we limit ourselves to the non-seismic constraints.

  We used the stellar evolution code ASTEC \citep[][2007]{JCD82b} to compute $\Xv(\thetav)$. The estimated parameters are the stellar mass, $M$, and age, $\stage$ ($\thetav=\{ M, \stage\}$). By neglecting diffusion and mixing processes other than convection, we can fix the metallicity ($Z=0.027$). The mixing-length parameter is assumed to be solar.

\begin{figure*}[t!]
\center
\resizebox{0.90\hsize}{!}{\includegraphics[clip=true]{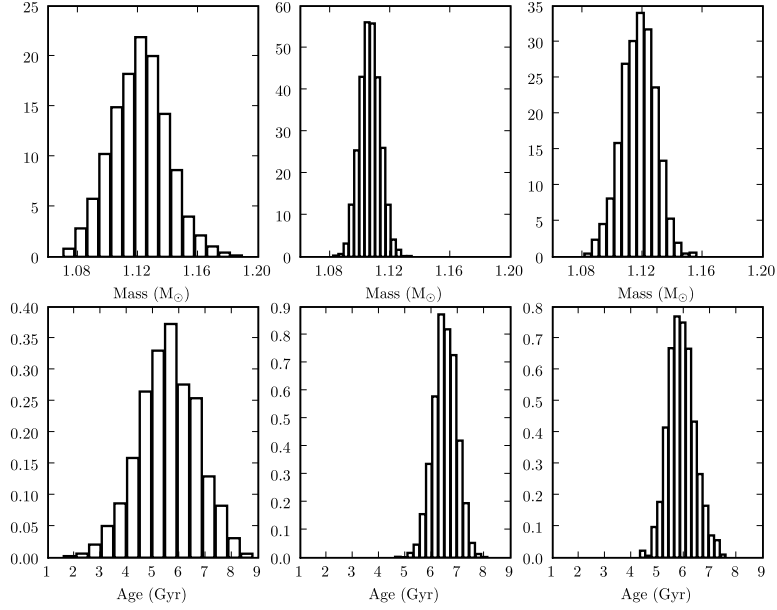}}
\caption{\footnotesize
 Marginal distributions obtained for the mass, $\pi(M|\Xv)$ (upper row), and the age, $\pi(\stage|\Xv)$ (lower row), of {\acena}. The columns correspond, from left to right, to runs 1 to 3 (see text for details).}
\label{marge}
\end{figure*}

  We present in Table~\ref{estim} and Fig.~\ref{marge}\ 
the results of three MCMC simulations using different observational constraints and priors. In run~1 we retained $\Xv= \{ L,T_{\mathrm{eff}}\}$ and uniform priors were used on $M$ and $\stage$. We used the same constraints in run~2 but the uniform prior on the mass was replaced by a Gaussian prior, $\pi(M) \propto \exp \left( - [M/M_{\odot}-1.105]^2/9.8\times 10^{-5} \right)$. Run~3 aims at testing the effect of precise radius measurements on the stellar parameters, we thus chose $\Xv = \{ L,T_{\mathrm{eff}},R \}$ and used uniform priors.

Table~\ref{estim} gives the inferred values for $M$ and $\stage$. We present the mean value and the corresponding standard deviation, derived from the marginal distributions $\pi(M|\Xv)$ and $\pi(\stage | \Xv)$ displayed in Fig.~\ref{marge}. The value estimated as the maximum {\it a posteriori} is also given. 
We can see immediately that there is a good agreement between the estimated values, which underlines the good agreement between independent observational constraints.

We observe a small discrepancy between the mean and MAP values obtained from run~1. This can be explained by the fact that convective cores start to appear in models with masses $\gtrsim 1.14$~{\msol}. 
 Non-linear effects associated with convective cores will, as a general trend, lead to a wider range of stellar parameters being able to reproduce the observations \citep[see, e.g.,][for an illustration of this phenomena]{Jorgensen05}, and hence to be accepted by the MCMC algorithm. Therefore the difference could be explained this way: by using the MAP value one selects only the model which reproduces best the observations, whereas the mean value accounts for all accepted models. It is worth noting that this discrepancy becomes extremely small for run~3 and almost disappears for run~2, during which no models with convective cores were accepted.

\bibliographystyle{aa}

\end{document}